\documentclass[aps,pra,showpacs,twocolumn,amsmath,amssymb]{revtex4}

\usepackage[utf8]{inputenc}
\usepackage{amssymb,amsthm,amsmath,amsfonts}
\usepackage{color, graphicx, enumerate}

\newtheorem{thm}{Theorem}

\newtheorem*{thm*}{Theorem}
\newtheorem*{lm*}{Lemma}
\newtheorem*{claim*}{Claim}

\theoremstyle{definition}

\newcommand{\R}{\mathbb{R}}

\newcommand{\Pb}[1]{\mathbb{P}\left( #1 \right)}

\newcommand{\dd}{\mathrm{d}}
\newcommand{\1}{\textbf{1}}

\newcommand{\p}[1]{\mathbb{P}\left( #1 \right)}

\DeclareMathOperator{\Var}{Var}

\DeclareMathOperator{\CPE}{CPE}
\DeclareMathOperator{\CUE}{CUE}
\DeclareMathOperator{\COE}{COE}

\begin{document}
\title{Extremal spacings between eigenphases of random unitary matrices \\ and their tensor products}

\author{Marek Smaczy{\'n}ski}
\affiliation{Smoluchowski Institute of Physics, Jagiellonian University, Reymonta 4, 30-059 Cracow, Poland.}

\author{Tomasz Tkocz}
\affiliation{Mathematics Institute, University of Warwick, Coventry CV4 7AL, UK.}

\author{Marek Ku{\'s}}
\affiliation{Center of Theoretical Physics, Polish Academy of Sciences, Al. Lotnik\'ow 32/46, 02-668 Warsaw, Poland.}

\author{Karol {\.Z}yczkowski}
\affiliation{Center of Theoretical Physics, Polish Academy of Sciences, Al. Lotnik\'ow 32/46, 02-668 Warsaw, Poland}
\affiliation{Smoluchowski Institute of Physics, Jagiellonian University, Reymonta 4, 30-059 Cracow, Poland.}

\date{\today}

\begin{abstract}
Extremal spacings between eigenphases of random unitary matrices of size $N$
pertaining to circular ensembles are investigated. Explicit probability
distributions for the minimal spacing for various ensembles are derived for
$N=4$.
We study ensembles of tensor product of $k$ random unitary matrices of size $n$
which describe independent evolution of a composite quantum system
consisting of $k$ subsystems. In the asymptotic case, as
the total dimension $N=n^k$ becomes large,
the nearest neighbor distribution $P(s)$ becomes Poissonian,
but statistics of extreme spacings $P(s_{\rm min})$ and $P(s_{\rm max})$
reveal certain deviations from the Poissonian behavior.
\end{abstract}


\maketitle

\section{Introduction}\label{sec:introduction}

Random unitary matrices are useful to describe spectra of periodic quantum
systems, the classical analogues of which are chaotic \cite{H06,S99}. The
choice of a specific ensemble of matrices is dictated by symmetry properties
of the investigated physical system. If the system possesses no time-reversal
symmetry the {\sl circular unitary ensemble} ($\CUE$) of matrices distributed
according to the Haar measure of the unitary group is appropriate
~\cite{Me04}. For systems with a generalized time reversal symmetry the {\sl
circular orthogonal ensemble} ($\COE$) describes properly statistical
properties of spectra if we neglect additional subtleties caused by specific
rotational symmetry features of systems with half-integer spin, which are of
no concern for investigations reported in this paper. In the case of
classically regular dynamics the spectrum of the evolution operator displays
level clustering characteristic to the {\sl circular Poissonian ensemble}
($\CPE$) of diagonal random unitary matrices. To describe intermediate
statistics one uses interpolating ensembles of unitary matrices
\cite{PS91,LZ92,ZK96} or composed ensembles of unitary matrices \cite{Poz}.
In the case of emerging chaos, in which the chaotic layer covers only a
fraction of the phase space of the classical system one may apply the
distribution of Berry and Robnik, originally used for autonomous systems
\cite{BR84}.

To characterize statistical properties of spectra of a random matrix one
often uses the nearest neighbor spacings  distribution $P(s)$
\cite{Me04,Fo10}. The random variable $s$ is the distance between adjacent
eigenphases (phases of eigenvalues) normalized by assuming that the mean
spacing is equal to unity.

In this work we investigate the distribution of yet another random variable
-- the minimal spacing $s_{\rm min}$ between two eigenphases. In similarity
to the standard statistics of nearest level spacings, also  the distribution
$P(s_{\min})$ encodes information about properties of the spectrum. Observe
that for any unitary matrix $U$ the size of its minimal spacing $s_{\rm min}$
provides an information, to which extent the investigated matrix $U$ is close
to be degenerated. For completeness we are also going to study the size of
the largest spacing $s_{\rm max}$ defined analogously.

Statistics of the minimal spacings in spectra of random Hermitian matrices
was analyzed by Ca{\"e}r et al. \cite{CMD07} and also discussed in the book
by Forrester \cite{Fo10}. Our current approach is somewhat similar but
different, as we investigate extremal gaps between eigenvalues of unitary
matrices distributed along the unit circle and study tensor products of
unitary matrices. After a part of our project was completed we learned about
a relevant work of Arous and Bourgade \cite{AB10} in which the distribution
of extremal spacings was studied for random  matrices of circular unitary
ensemble.

The paper is organized as follows. For exemplary ensembles of random matrices
of size $N=4$ we derive in Section~\ref{sec:case_study} exact forms of the
distributions of minimal spacings. The chosen dimension allows exact
calculations, which become rather complicated for larger matrices. Moreover,
this is the minimal dimension in which results for $\CUE$ and $\CPE$ can be
compared with those for the ensemble consisting of tensor products of two
$\CUE$ random matrices of size $N=2$. Such an ensemble corresponds to a
generic local dynamics in a two-qubit system~\cite{TSKZZ12}.

The case of large matrices is studied in
Section~\ref{sec:extremal_statistics}. We recall the heuristic argument put
forward e.g. in  \cite{Fo10} (see Exercise 14.6.5, p. 697) justifying that
for a random unitary matrix of size $N$ the size of the minimal gap scales as
$ s_{\min} \approx N^{-\frac{1}{1+\beta}}$ where $\beta=0,1$ and $2$ for the
Poissonian, orthogonal and unitary circular ensemble, respectively.
Analogously, we approach the asymptotic scaling of the maximal gap
$s_{\max}$. We also provide some numerical results confirming our
non-rigorous predictions concerning the order of the mean values of the
extremal spacings $\langle s_{\min} \rangle$, $\langle s_{\max} \rangle$, and
the distribution of the minimal spacing $s_{\min}$. Recently, the latter has
been rigourously studied in \cite{AB10} and \cite{Sh07}. It was considered
for the first time in \cite{Vi01}.

Furthermore, we analyze extremal spacings for products of $k$ independent
random unitary matrices, each of them of size $n$. If the total dimension of
the matrix, $N=n^k$, is large the level spacing distribution $P(s)$ becomes
Poissonian \cite{TSKZZ12}. This property holds also for a tensor product of
two random unitary matrices of a different size \cite{Tk13}. However, in the
case of a large number of one-qubit systems, ($n=2$ and $k$ large),
statistics of the minimal spacing $s_{\rm min}$ displays significant
deviations from the predictions for the Poisson ensemble, reviewed in the
Appendix.

We use the following notation. For a single unitary or orthogonal matrix $A$
of size $N$ we consider its spectrum $\{\exp(i\varphi_j)\}_{j=1}^N$, where
$(\varphi_1, \ldots, \varphi_N)$ represents the vector of the eigenphases
ordered non-decreasingly, $0 \leq \varphi_1 \leq \ldots \leq \varphi_N <
2\pi$. We order non-decreasingly the spacings $\varphi_2 - \varphi_1, \ldots,
\varphi_N - \varphi_{N-1}, 2\pi + \varphi_1 - \varphi_N$ between neighboring
eigenphases, divide them by the average spacing $2 \pi /N$
 and denote the obtained sequence by
\begin{equation}\label{eq:def}
	s_{\rm min} := s_1 \leq \ldots \leq s_N =: s_{\rm max}.
\end{equation}
The standard level spacing distribution
$P(s)$ is given by the average  $\frac{1}{N}\sum_{m=1}^N P_m(s_m)$,
 where $P_m$ is the density of the rescaled $m$-th spacing
 $s_m=(\varphi_{m+1} - \varphi_m)N/2\pi$.

\section{Case study:  minimal spacings for two--qubit system}
\label{sec:case_study}

Our first goal is to derive  exact probability distributions of the minimal
spacing $P_{\min}$ for exemplary ensembles of random unitary matrices of size
$N=4$. Besides the Poissonian and the unitary ensemble we analyze also the
tensor product of two independent random matrices of size $N=2$. This
ensemble, denoted for brevity as  {\sl CU}$E_{2 \otimes 2}$, describes
dynamics of two independent quantum sub-systems \cite{TSKZZ12}. In the
quantum information literature such a case is called a two--qubit system.

To derive the desired distribution we calculate the tail distribution
$T(t) = \Pb{s_{\min} > t}$ and take the derivative of $T$.
We have
\begin{equation}\label{eq2:1}
\begin{split}
	T(t) &= \Pb{s_{\min} > t} \\ &= \Pb{\varphi_2 - \varphi_1, \varphi_3 - \varphi_2, \varphi_4 - \varphi_3, 2\pi + \varphi_1 - \varphi_4 > \pi t/2} \\
	&= \int_{\{\varphi_2 - \varphi_1, \varphi_3 - \varphi_2, \varphi_4 - \varphi_3, 2\pi + \varphi_1 - \varphi_4 > \pi t/2\}} \\
	&P^{\textrm{ord}}\left( \varphi_1, \varphi_2, \varphi_3, \varphi_4 \right)d\left( \varphi_1, \varphi_2, \varphi_3, \varphi_4 \right),
\end{split}
\end{equation}
where $P^{\textrm{ord}}$ is the joint probability distribution of ordered
eigenphases, which can be obtained from the joint probability distribution
for a given ensemble. After changing variables,
$\psi_1 = \varphi_1,  \psi_2 = \varphi_2 - \varphi_1,
  \psi_3 = \varphi_3 - \varphi_2$ and  $\psi_4 = \varphi_4 - \varphi_3$,
the integration domain splits into two tetrahedrons.
%
Standard  but tedious calculations yield in each case the tail distribution
function $T(t)$,
which leads to the corresponding
probability density,  $P(s_{\min}) =-\frac{\dd}{\dd t}T(t) |_{t=s_{\min}}$.

\begin{enumerate}[(a)]
   \item for $\CUE_{2 \otimes 2}$,
   \begin{equation}
\label{eq2:Smincuexcue}
   \begin{split}
      P^U_{2 \otimes 2}(s_{\min}) = \frac{1}{4\pi}\Big(&2\pi (1 - s_{\min})\big(4 - \cos (\pi s_{\min}/2)\big) \\ &- 3\sin(\pi s_{\min}/2)
     + 8\sin(\pi s_{\min}) \\  &- 3\sin(3\pi s_{\min}/2)\Big),
   \end{split}
   \end{equation}
   \item for $\CUE_4$
	\begin{equation}\label{eq2:Smincue}
	\begin{split}
      P^U_4(s_{\min}) = &\frac{1}{72\pi^2}\sin^2(\pi s_{\min}/4)\bigg(666 + 720\pi^2(1 - s_{\min})^2 \\
      &+ 36 \big(11 + 16\pi^2 (1 - s_{\min})^2\big) \cos(\pi s_{\min}/2) \\
      &+ 18\big(8\pi^2 (1 - s_{\min})^2 - 13\big)\cos(\pi s_{\min})\\
      &- 100\cos(3 \pi s_{\min}/2) - 608\cos(2 \pi s_{\min}) \\
      &- 380\cos(5 \pi s_{\min}/2) + 234\cos(3 \pi s_{\min}) \\
      &+ 74\cos(7 \pi s_{\min}/2) - 58\cos(4 \pi s_{\min}) \\
      &+ 10\cos(9 \pi s_{\min}/2) \\
      &+ 24\pi(1 - s_{\min})\Big[60\sin(\pi s_{\min}/2) \\
      &+ 63\sin(\pi s_{\min}) + 22\sin(3 \pi s_{\min}/2)\\
      &+ 2\sin(2\pi s_{\min}) - 4\sin(5 \pi s_{\min}/2)\Big] \bigg),
	\end{split}
   \end{equation}
   \item for $\CPE_4$
   \begin{equation}\label{eq2:Smincpe}
      P^P_4(s_{\min}) = 3(1 - s_{\min})^2.
   \end{equation}
\end{enumerate}

\begin{figure}[ht!]
  \begin{center}
    \scalebox{0.6}{\includegraphics[width=0.8\textwidth]{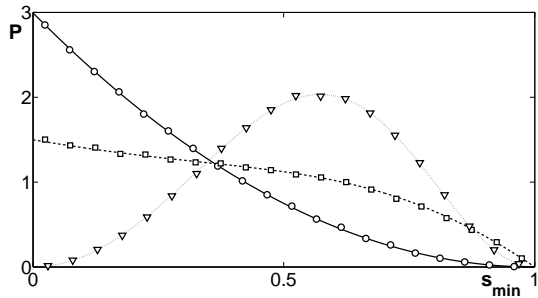}}
   	\caption{Probability densities of the minimal spacing $s_{\min}$
for random matrices
   of size $N=4$ pertaining to $\CUE_4$  ($\triangledown$),
 $\CUE_{2 \otimes 2}$ ($\square$), and   $\CPE_4$ ($\circ$).
Symbols denote numerical
   results obtained for $2^{14}$ independent matrices, while the curves
   represent distributions \eqref{eq2:Smincuexcue},
   \eqref{eq2:Smincue} and \eqref{eq2:Smincpe},
   respectively.\label{fig2:Smin}}
    \end{center}
\end{figure}


These three distributions are presented in Fig.~\ref{fig2:Smin}. The behavior
of the densities around zero encodes some information concerning level
repulsion and level clustering. The variable $s_{\min}$ is the smallest
distance between two neighboring eigenphases. Therefore, the fact that its
density is separated from zero, say $P(s_{min}) > 1$ for $s_{min}$ close to
zero, means that for a small $\epsilon > 0$ the probability that some two
phases are at the distance closer than $\epsilon$ equals $\Pb{s_{\min} <
\epsilon} = \int_0^\epsilon P(s_{min}) \dd s > \epsilon$. In the cases of
$\CPE_4$ and $\CUE_4$ these features are consistent with level clustering and
level repulsion observed in the distribution of spacings $P(s)$.
Fig.~\ref{fig2:Smin} shows that the eigenphases of the tensor product
$\CUE_{2 \otimes 2}$ tend to accumulate in a spectacular contrast to the case
of a single random unitary matrix form $\CUE$ \cite{TSKZZ12}.

Numerical results show that for large $N$
the distributions of the $m$--th spacing $P(s_m)$ are close
to the level spacing distribution $P(s)$ for $m \approx N/2$.
 However, for any $N$ the
distributions  of the smallest spacing $s_{\min}=s_1$ and of
the largest spacing $s_{\max}=s_N$ differ considerably. We shall then analyze these
distributions of extremal spacings, which can be used as auxiliary
statistical tools to characterize ensembles of random matrices.

\section{Extremal statistics for large matrices}
\label{sec:extremal_statistics}

In this section we analyze extremal gaps in the spectra of circular ensembles
of random matrices of a large size, $N\gg 1$, giving the numerical evidence
to support some simple heuristic arguments (the subject for $\CUE$ ensemble
has been rigorously studied though, see e.g. \cite{AB10}). As usual, we
parameterize canonical ensembles by the level repulsion parameter $\beta$,
equal  to $0,1$ and $2$ for Poissonian, orthogonal and unitary ensembles
respectively. The relevant quantities are labeled by the index $\beta = 0, 1,
2$. For instance $P_\beta(s)$ represents the level spacing distribution for
the corresponding ensemble of random unitary matrices. We shall start with
the Poissonian ensemble described by the case $\beta=0$. Some basic
properties of the {\sl Poissonian process} are reviewed in the Appendix A.

\subsection{Asymptotics of the extreme spacings for Poisson process}

We are interested in asymptotic properties of spectra of diagonal random
unitary matrices. We choose at random  $N$ points from the unit circle
 $\{z \in \mathbb{C}, |z| = 1\}$, each independently according to the uniform distribution.
The arguments of these points ordered non-decreasingly will be called
 $0 \leq \theta_1 \leq \ldots \leq \theta_N < 2\pi$.
We define a point process $\Xi_N$ of the
rescaled eigenphases of a diagonal random unitary matrix $D_N = \textrm{diag }(e^{i\theta_1}, \ldots, e^{i\theta_N})$
pertaining to CPE$_N$,
\begin{equation}\label{eq:defXi_N}
	\Xi_N = \{(N/2\pi)\theta_1, \ldots, (N/2\pi)\theta_N\}.
\end{equation}
Moreover, we define the spacings $s_i$, $s_{\min}$, and $s_{\max}$ according to \eqref{eq:def}. Note that the scaling is chosen so that the mean spacing $\langle s \rangle$ is fixed to unity.

For the standard Poisson process $\Pi = \{X_1, X_2, \ldots\}$  (see Appendix
A), where its points are labeled in the nondecreasing order $0 \leq X_1 \leq
X_2 \leq \ldots$, we also define the spacings
\begin{equation}\label{eq:defspacingsofPi}
	Y_1 = X_1, \ Y_2 = X_2 - X_1, \ Y_3 = X_3 - X_2, \ \ldots.
\end{equation}

It is known that
for large $N$ the process $\Xi_N$ becomes Poissonian,
as the correlation functions converge to the constant functions equal to unity
characteristic of the Poisson process $\Pi$.


We would like to address the question of the asymptotic behavior of the variables $s_{\min}$ and $s_{\max}$. Since for a diagonal unitary matrix of CPE
the process (\ref{eq:defXi_N}) becomes Poissonian, the variables $\min_{j \leq N} Y_j$ and $\max_{j \leq N} Y_j$ satisfy
\begin{equation}\label{eq:linkbetweenSandY}
\begin{split}
	\sup_{t \in \R} \left| \p{s_{\min} \leq t} - \p{\min_{j \leq N} Y_j
 \leq t} \right| &\xrightarrow[N\to\infty]{} 0, \\
	\sup_{t \in \R} \left| \p{s_{\max} \leq t} - \p{\max_{j \leq N} Y_j
 \leq t} \right| &\xrightarrow[N\to\infty]{} 0.
\end{split}
\end{equation}

In view of \eqref{eq:linkbetweenSandY} we arrive at the desired
 conclusions regarding $s_{\min}$ and $s_{\max}$. These quantities are of order
\begin{equation}\label{eq:meansminmax}
	{\langle s_{\min} \rangle}_{\CPE} \sim 1/N, \quad {\langle s_{\max} \rangle}_{\CPE} \sim \ln N.
\end{equation}

After rescaling $s_{\min}$ converges to a random variable $y$ with
exponential density,
\begin{equation}\label{eq:sconvergence}
	Ns_{\min} \overset{d}{\longrightarrow} e^{-y}\1_{\{y > 0\}},
\end{equation}
where by $\1_Y$ we denote the characteristic function of the set $Y$. The
maximal spacing $s_{\max}$ converges to a constant,
\begin{equation}\label{eq:sconvergence1}
      s_{\max}/\ln N \overset{d}{\longrightarrow} 1,
\end{equation}
where $\overset{d}{\longrightarrow}$ denotes the convergence in distribution.

The fluctuations of the rescaled variable $s_{\max}/\ln N$ around $1$ are of order $1/\ln N$ and they are described by the Gumbel distribution,
\begin{equation}\label{eq:sfluctuations}
	s_{\max} - \langle s_{\max} \rangle \overset{d}{\longrightarrow} P(x)\sim e^{-(x+\gamma)-e^{-(x+\gamma)}}.
\end{equation}
Here and throughout, we denote by $\gamma\approx 0.5772$ Euler's constant.


\subsection{Mean minimal spacing}\label{sec:extremal_statistics:<S_min>}


For the sake of convenience, we recall here the heurisitic reasoning leading to the estimate of mean of the minimal gap (Exercise 14.6.5 in \cite{Fo10}). In the next subsection we follow this idea to deal with the maximal gap.

To get an estimation of the behavior of the mean minimal spacing
of a random unitary matrix of size $N$ let us assume that spacings $s_j$, $j=1,\dots, N$ are independent
random variables. For small spacing one has $P_{\beta}(s) \sim s^{\beta}$,
so the integrated distribution $I_\beta(s)=\int_0^s P_\beta(s')ds'$ behaves as
$I_{\beta}(s) \sim s^{1+\beta}$. A matrix $U$ of size $N$ yields $N$ spacings
$s_j$. Thus the minimal spacing $s_{\min}$ occurs on average for such an
argument of the integrated distribution that $I_{\beta}(s_{\min}) \approx
1/N$. This implies that $(s_{\min})^{1+\beta} \approx 1/N,$ which allows us
to estimate the average minimal spacing
\begin{equation}
\label{eq1:average_Smin}
  \langle s_{\min} \rangle \approx N^{-\frac{1}{1+\beta}}.
\end{equation}

\begin{figure}[ht!]
  \begin{center}
    \scalebox{0.7}{\includegraphics[width=0.7\textwidth]
{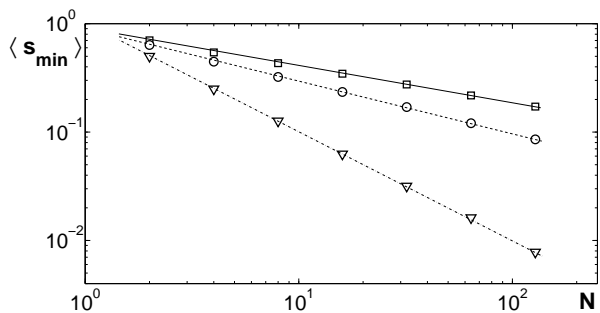}}
   	\caption{Mean minimal spacing $\langle s_{\min}\rangle$
  as a function of the matrix size $N=2^m$ for ($\triangledown$) $\CPE$,
 ($\square$) $\COE$   and ($\circ$) $\CUE$ and $m=1,...,7$.
 Symbols denote numerical results obtained for $2^{14}$ independent random
 matrices. Solid, dashed and dash-dot lines are plotted with
 slopes implied by the estimation \eqref{eq1:average_Smin}
 and equal to $-1$, $-1/2$ and  $-1/3$, respectively.
 Linear fit to numerical data yields slopes -0.98, -0.48, -0.33, respectively.}
 \label{fig3:fig_avmin}   \end{center}
\end{figure}

In the case $\beta=2$ corresponding to $\CUE$ this statement is consistent
with the rigorous results \cite{AB10} of Arous and Bourgade. As shown in
Fig.~\ref{fig3:fig_avmin} the above heuristic reasoning provides the correct
value of the exponent in dependence of the mean minimal spacing $\langle
s_{\min} \rangle$ on the matrix size $N$ for $\CPE$ $(\beta=0)$, $\COE$
$(\beta=1)$ and $\CUE$ $(\beta=2)$.

\subsection{Mean maximal spacing}\label{sec:extremal_statistics:<S_max>}

\begin{figure}[ht!]
  \begin{center}
    \scalebox{1.0}{\includegraphics[width=0.4\textwidth]{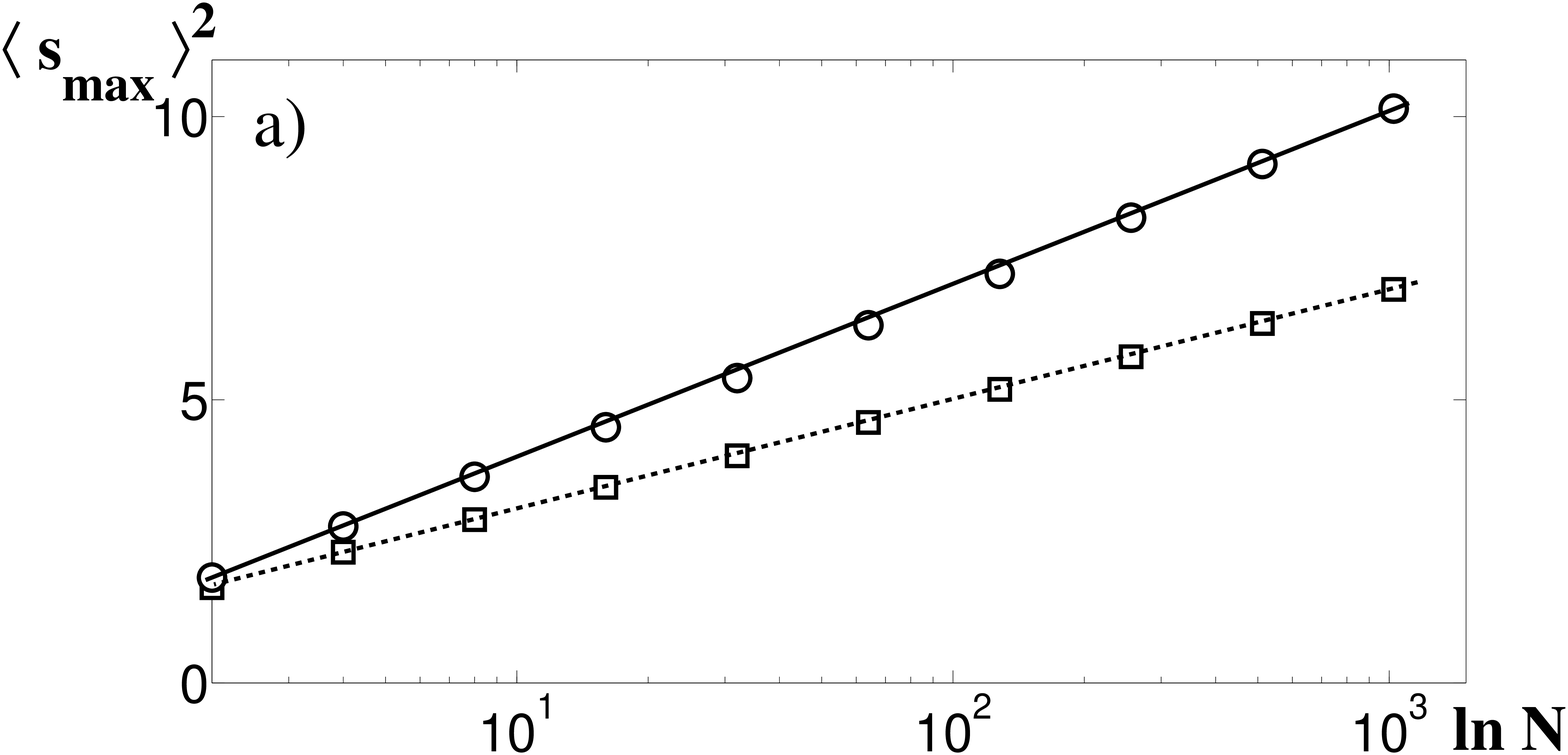} }
\scalebox{1.0}{\includegraphics[width=0.4\textwidth]{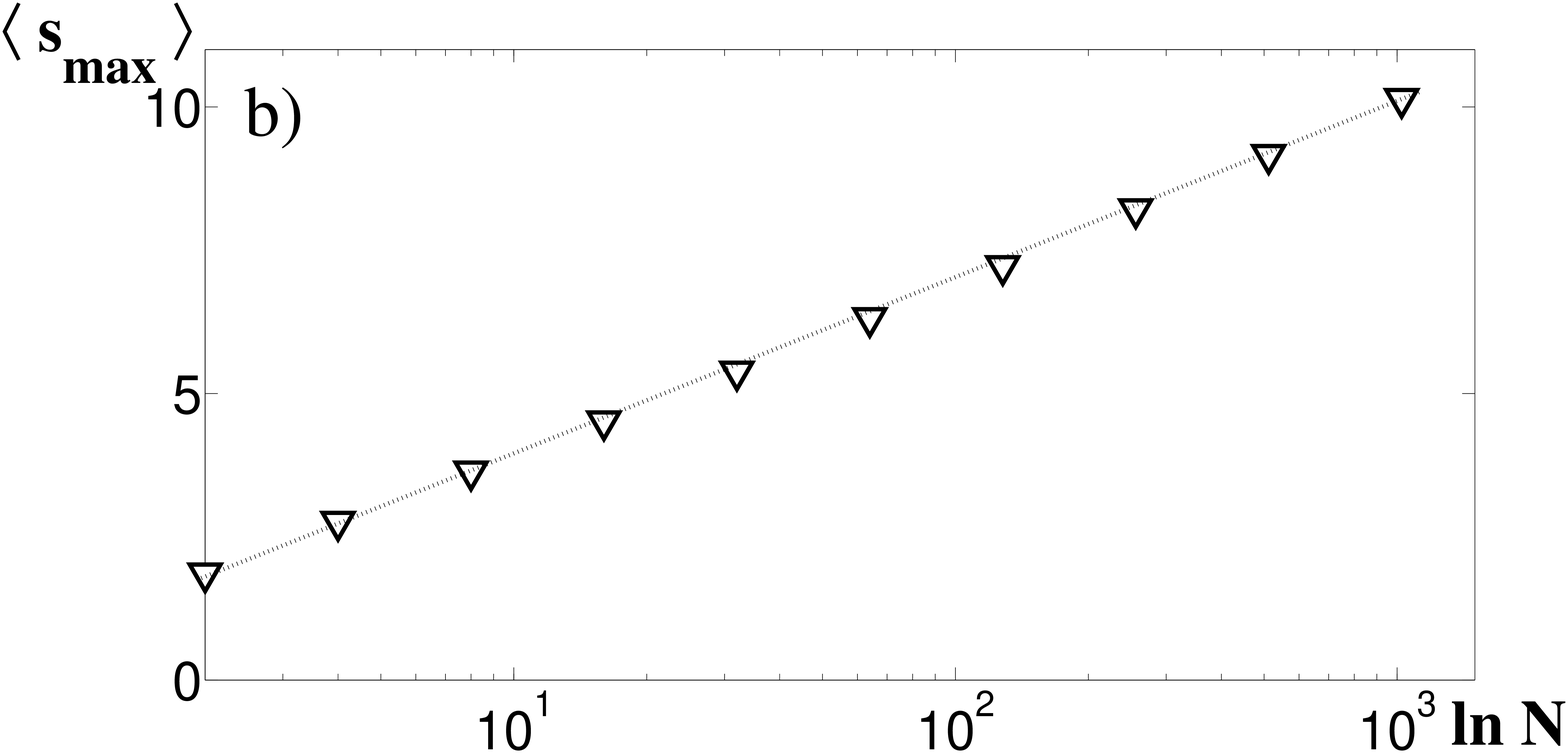} }
 \caption{Mean maximal spacing $\langle s_{\max}\rangle$ as a function of the
 matrix size $N=2^k$ with $k=1,...,10$ plotted for a)
 $\CUE$ ($\circ$), $\beta=2$;  $\COE$ ($\square$), $\beta=1$;
 and  b)  $\CPE$ ($\triangledown$), $\beta=0$.
 Symbols denote numerical results
 obtained for $2^{14}$ independent random matrices. Solid, dashed (panel a) and
 dash-dot (panel b) lines are plotted with slopes
implied by estimations (\ref{avsmaxcoe}),
 (\ref{avsmaxcue}) and (\ref{avsmaxcpe}), respectively.}
\label{fig3.2}
  \end{center}
\end{figure}

We study the average maximal spacing $\langle s_{\max} \rangle$ for random unitary matrices of the circular orthogonal ensemble.
Matrix of size $N$ yields $N$ spacings $s_j$. In analogy to the previous reasoning we shall assume that all
spacings are independent random variables described by the Wigner surmise
\begin{align}
 P(s) = \frac{\pi}{2} s e^{ - \pi s^2/4}.
\label{Wigner_coe}
\end{align}
Thus the integrated distribution $I(s)=\int_0^s P(s')ds'$
reads  $I(s) = 1-e^{-\pi s^2/4}$.
The maximal spacing $s_{\max}$ occurs on average for such an argument
of the integrated distribution function
that $1-I(s_{\max}) \approx 1/N.$
This implies that $e^{ - \pi s_{\max}^2/4} \approx 1/N$, which
 allows us to estimate the average maximal spacing,
\begin{equation}
  \langle s_{\max} \rangle^2_{COE} \approx \frac{4}{\pi} \ln N.
\label{avsmaxcoe}
\end{equation}
This implies that $\langle s_{\max}\rangle^2$ grows with the matrix size N
proportionally to $\frac{4}{\pi} \ln N$ what is demonstrated in Fig. (\ref{fig3.2}).

\smallskip

Let us deal now with the circular unitary ensemble. We employ here the Wigner
formula for the level spacing distribution of a large $\CUE$ matrix,
 $P_2(s) = \frac{32}{\pi^2}s^2e^{-4s^2/\pi}$. By the
same reasoning as above we obtain an estimate that the maximal spacing
$s_{\max}$ occurs on average for such an argument of the integrated
distribution function
 $I(s) = \int_0^s P(s')ds'$ that $1-I(s_{\max}) \approx 1/N$. Thus
\begin{equation}
	\frac{1}{N} \approx \int_{s_{\max}}^\infty \frac{32}{\pi^2}s^2e^{-4s^2/\pi} \dd s.
\end{equation}
We change the variable setting $u = 4s^2/\pi$ and obtain $\frac{1}{N} \approx \int_{4s_{\max}^2/\pi}^\infty \frac{2}{\sqrt{\pi}}u^{1/2}e^{-u} \dd u. $
Therefore, supposing $s_{\max}$ is large we get
\begin{equation}
	\frac{1}{N} \approx \frac{4}{\pi}s_{\max} e^{-4s_{\max}^2/\pi}.
\end{equation}
Now we take the logarithm of both sides, neglect $\ln s_{\max}$ as it is of lower order than $s_{\max}^2$ for large $s_{\max}$, and arrive at
\begin{equation}
	\langle s_{\max} \rangle^2_{CUE} \approx \frac{\pi}{4} \ln N .
\label{avsmaxcue}
\end{equation}

\medskip

In the case of a Poissonian spectrum the level spacing distribution displays
an exponential tail, $P(s) \sim e^{-s}$. Thus the integrated
distribution function $I(s)=\int_0^s P(s')ds'$ behaves as $I(s) =1-e^{-s}$.
For a matrix of size $N$ the maximal spacing $s_{\max}$ occurs on average for
such an argument that $1-I(s_{\max}) \approx 1/N$. This implies that
$e^{-s_{\max}}
 \approx 1/N$ and enables us to estimate the average maximal spacing for the circular Poisson ensemble as
\begin{equation}
  \langle  s_{\max} \rangle_{CPE} \approx \ln N.
\label{avsmaxcpe}
\end{equation}
Analyzing estimations following from eqn. (\ref{avsmaxcoe}), (\ref{avsmaxcue}) and (\ref{avsmaxcpe})
one obtains slopes $A_{COE}=\frac{4}{\pi} \approx 1.27$, $A_{CUE}=\frac{\pi}{4} \approx 0.77$
and $A_{CPE}=1$, which are comparable with numerical results $A_{COE} \approx 1.33$, $A_{CUE} \approx 0.84$
 and $A_{CPE} \approx 0.97$, presented in Fig. \ref{fig3.2}.

\subsection{Distribution of extremal spacings}
\label{sec:extremal_statistics:P(S_k)}

To study the distributions of the minimal spacing $s_{\rm min}$ we introduce
a rescaled variable suggested by (\ref{eq1:average_Smin}),
\begin{equation}
\label{eq:x_beta}
 x^{(\beta)}_{\min}:=A_{(\beta)} N^{\frac{1}{1+\beta}}s_{\min},
\end{equation}
where $A_{(\beta)}$ is a constant, in general different for $\CPE$, $\COE$
and $\CUE$.

\begin{figure}[ht!]
  \begin{center}
    \scalebox{0.5}{\includegraphics[width=1.0\textwidth]{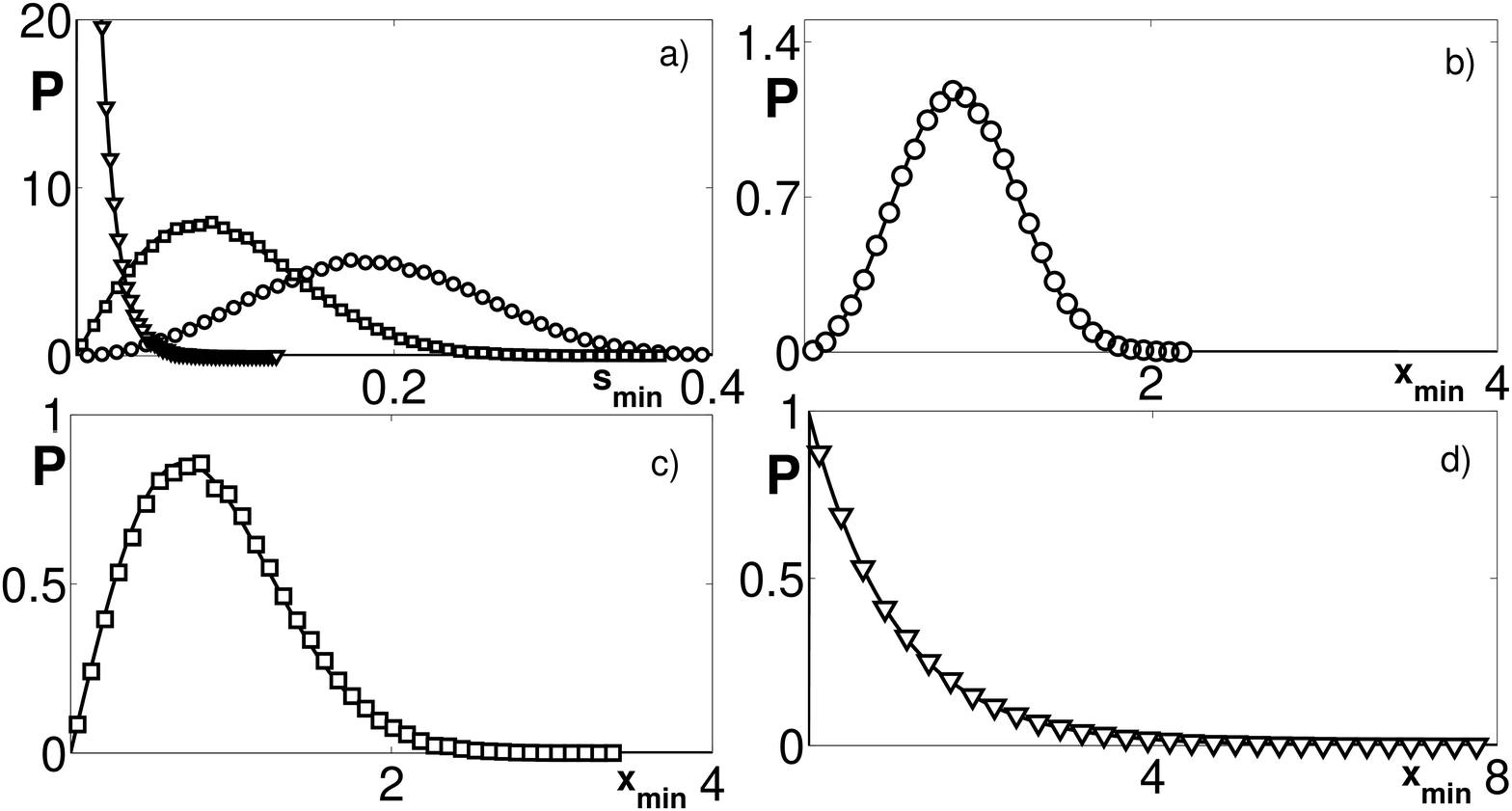}}
   \caption{Probability distributions a) $P(s_{min})$
for random unitary matrices of  $\CUE$ ($\triangledown$), $\beta=2$;
 $\COE$  ($\square$), $\beta=1$;  and  $\CPE$ ($\circ$), $\beta=0$.
The same data shown for variable $x_{\rm min}$ rescaled according to
(\ref{eq:x_beta}) for b) $\CUE$, c) $\COE$ and d) $\CPE$.
Symbols denote numerical results obtained
for $2^{17}$ independent matrices of size $N=100$,
while solid curves represent asymptotic predictions  (\ref{pxx}).}
\label{fig_psmin}
   \end{center}
\end{figure}

The case of the unitary ensemble was recently studied by Arous and Bourgade
\cite{AB10}, who  derived the following expression for the asymptotic
distribution of the minimal spacing,
\begin{equation}
P(x_{\min})=3x_{\min}^2 e^{-x_{\min}^3},
\label{distmincue}
\end{equation}
in the rescaled variable $x_{\min}=(\pi/3)^{2/3}N^{1/3}s_{\min}$. This result
suggests the following general form of the distribution of minimal spacing
for all three ensembles considered labeled by the level repulsion parameter
$\beta$,
\begin{equation}
\label{pxx}
 P^{(\beta)}(x_{\min}):=(\beta+1) x^{\beta}_{\min} e^{-x^{\beta+1}_{\min}},
\end{equation}
which agrees with the numerical data -- see Fig.~\ref{fig_psmin}.

The above formula has a structure  $F(x):=\frac{df(x)}{dx}e^{-f(x)}$, which
helps to determine the normalization. Numerical results suggest that
constants read: $A_{(0)}=1$ for $\CPE$,  $A_{(1)}=\langle s \rangle$ for
$\COE$, and $A_{(2)}=(\pi/3)^{2/3}$ for $\CUE$.

Returning to the original variable $s_{\min}$ we obtain the distributions
$P^{(\beta)}(s_{\min})$,
\begin{equation}
 P^{(0)}(s_{\min})=A_{(0)} N e^{-N s_{\min}},
\end{equation}
\begin{equation}
 P^{(1)}(s_{\min})=2 A_{(1)}^2 N s_{\min} e^{-A_{(1)}^2 N s_{\min}^2},
\end{equation}
\begin{equation}
 P^{(2)}(s_{\min})=3 A_{(2)}^3 N s_{\min}^2 e^{-A_{(2)}^3 N s_{\min}^3}.
\end{equation}

The distributions of the minimal spacing obtained numerically for Poisson,
 orthogonal and unitary circular ensembles of random matrices of the size $N=100$ are presented in Fig.~\ref{fig_psmin}.

\section{Extremal spacings for tensor products of random unitary matrices}

In this section we study eigenphases of tensor products of random unitary matrices.
 We are interested in two cases
\begin{enumerate}[A)]
	\item\label{case1} Two--qunit system: Given two independent $\CUE$
matrices $U_A, U_B$ of size $n$ with eigenphases $\{\psi_j\}_{j=1}^n$,
$\{\phi_j\}_{j=1}^n$ respectively, define the point process $\Xi_n$ of
the rescaled eigenphases of the tensor product $U_A \otimes U_B$
	\begin{equation}\label{eq:defAotimesBpointprocess}
\Xi_n = (n^2/2\pi)\left\{ (\psi_i + \phi_j) \mod 2\pi, \ i,j = 1, \ldots, n
 \right\}.
	\end{equation}
	\item\label{case2} $k$--qubit system: Given $k$ independent $\CUE$
matrices of order two, $V_1, \ldots, V_k$ with eigenphases $\{\psi_{m,1},
\psi_{m,2}\}$, $m = 1,\ldots, k$ respectively, define the point process
$\Psi_k$ of the rescaled eigenphases of the tensor product $V_1 \otimes
\ldots \otimes V_k$
	\begin{equation}\label{eq:defAotimes^npointprocess}
		\Psi_k = (2^k/2\pi)\left\{ \sum_{m=1}^k \psi_{m,\epsilon_m}
     \mod 2\pi, \ \epsilon_k, \ldots, \epsilon_k \in \{1,2\} \right\}.
	\end{equation}
\end{enumerate}
It has been recently shown
that both the process $\Xi_n$ and $\Psi_k$ asymptotically behave as the standard Poisson point process $\Pi$
-- see \cite{TSKZZ12} and Appendix A.
 Therefore, one might expect that the extremal spacings of the processes $\Xi_n$
 and $\Psi_k$ also exhibit the asymptotic of the extremal spacings of the
 Poisson process $\Pi$.

We have studied the problem numerically. To investigate the asymptotic regime
we analyzed large matrices, which cannot be diagonalized directly. In case
\ref{case2}), for instance, to deal with a $20$--qubit system one has to work
with matrices of size $N=2^{20} > 10^6$. To obtain eigenphases and, in
consequence, the desired distribution of level spacings, we adopted another
strategy summarized in the following algorithm.

1. Take an ensemble of $k$ random unitary matrices  $U_j$ of size two distributed according to the Haar measure \cite{ZK96,MEZ}.

2. Diagonalize them to obtain their spectra, $\{e^{i \varphi_{jm}}\}$, where $j=1,\dots,k$
labels  the number of the matrix, while $m=1,2$ labels eigenvalues of the $j$-th matrix.

3. Construct $N=2^k$ eigenphases of the tensor product $U=U_1\otimes \cdots \otimes U_k$,
by summing all combinations of phases from different matrices,
$\psi_{m_1,\dots m_k}=\sum_{j=1}^k \varphi_{jm_j}|_{{\rm mod} 2\pi},$ where $m_j=1,2$.

4. Order nondecreasingly the spectrum of $U$ containing $N=2^k$ eigenphases,
 $0 \leq \psi_1 \leq \ldots \leq \psi_{N} \leq 2\pi$.

5. Compute spacings between neighboring eigenphases, $s_1=(\psi_2 - \psi_1)N/2\pi, \ldots, s_{N-1}=(\psi_N - \psi_{N-1})N/2\pi, s_{N}=(2\pi + \psi_1 - \psi_N)N/2\pi$,
 order them nondecreasingly,
 find the minimal spacing $s_{\rm min}$ and the maximal spacing $s_{max}$.

\begin{figure}[ht!]
  \begin{center}
 \scalebox{0.5}{\includegraphics[width=0.6\textwidth]{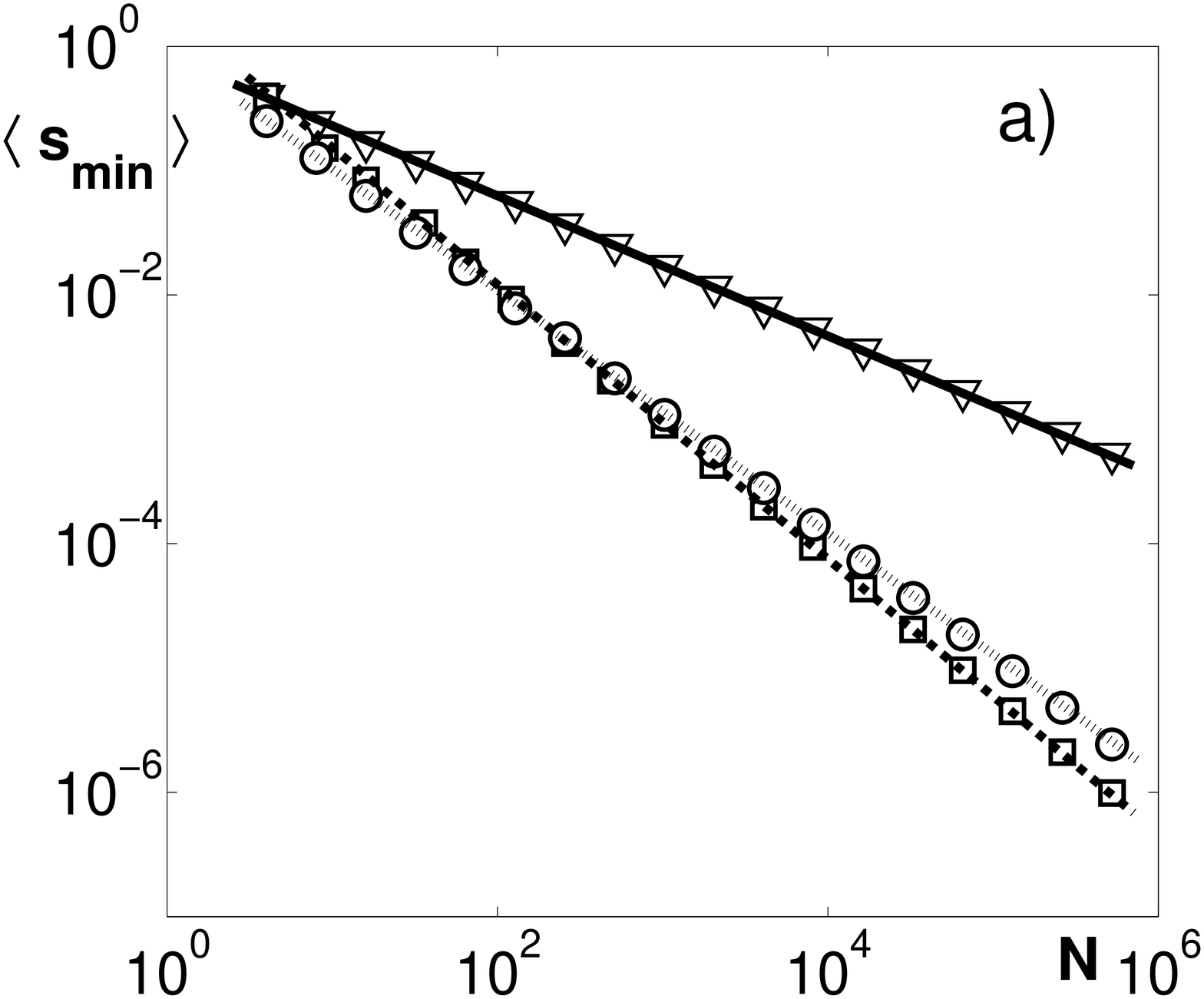}}
\scalebox{0.5}{\includegraphics[width=0.6\textwidth]{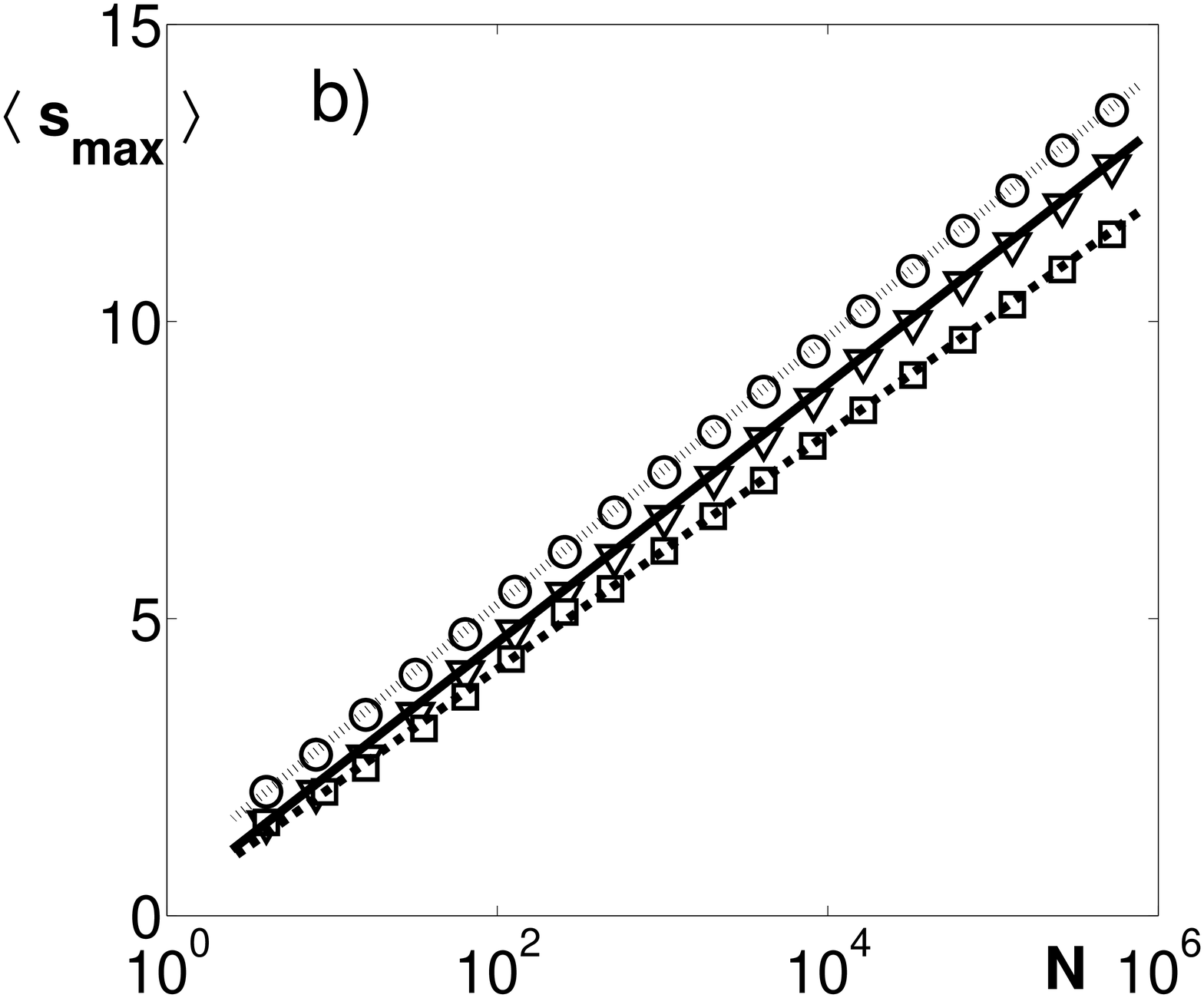}}
 \caption{Dependence of the mean extremal spacing on the matrix size $N$.
 a) Mean minimal spacing $\langle s_{\min}\rangle$,
assumed to behave as $N^{\eta}$ is plotted in log--log scale,
and the fitted exponents read
 $\eta_{(P)}=-0.98$ for $\CPE_N$ ($\circ$) ,
 $\eta_{(2)}=-1.09$ for $\CUE_{n \otimes n}$ ($\square$) ,
 $\eta_{(k)}=-0.58$ for $\CUE_2^{ \otimes k}$ ($\triangledown$).
  b) Mean maximal spacing  $\langle s_{\max}\rangle$
assumed to behave as ${\kappa \log N}$  and  plotted in log--linear scale,
with fitted prefactors $\kappa_{(P)}=0.98$ for $\CPE_N$ ($\circ$) ,
 $\kappa_{(2)}=0.85$ for $\CUE_{n \otimes n}$ ($\square$) ,
 $\kappa_{(k)}=0.95$ for $\CUE_2^{ \otimes k}$ ($\triangledown$).
 Symbols denote numerical results obtained for $2^{14}$ independent random matrices.
Solid, dashed and dash-dot represent the fitted lines.}
  \label{fig4:All}\end{center}
\end{figure}

Such a procedure allowed us to achieve $N$ above $10^6$ with a minor
numerical effort - see Fig.~\ref{fig4:All}. A similar procedure was be used
in case \ref{case1}) corresponding to the two--qunit system. Taking two
independent random unitary matrices $U_1$ and $U_2$ of size $n=1000$
diagonalizing them and adding the phases modulo $2\pi$ we constructing the
spectrum of the tensor product, $U=U_1\otimes  U_2$ of size $n^2$. In this
way we computed averages taken over the ensemble of tensor product matrices
of order $N=10^6$.

Dependence of the mean extremal spacings on the matrix size $N$ for tensor
products of case A) (two-qunits) and case B) ($k$--qubits) are shown in
Fig.\ref{fig4:All}. Panel a) shows the average {\sl minimal} spacing $\langle
s_{\rm min}\rangle$. Note that the scaling of the minimal spacing for the two
subsystems of size $n$ ($\square$) agrees with the Poissonian predictions. On
the other hand, in the case of the system consisting of $k$ qubits, the
scaling exponent is close to $-0.6$ and differs considerably from the value
$-1$ characteristic to the Poissonian ensemble. As shown in
Fig.~\ref{fig4:All}b, behavior of the average maximal spacing for the tensor
products corresponding to $N=n \times n$ and $N=2^k$ systems is closer to the
prediction of the Poisson ensemble, $\langle s_{\rm max}\rangle \sim \ln N$.

\subsection{Minimal spacings for tensor products}



\begin{figure}[htp]
	\centering
	\includegraphics[width=0.5\textwidth]{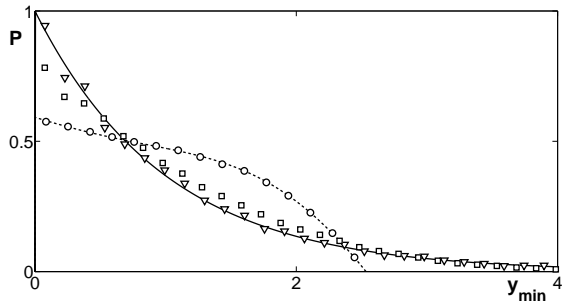}
	\caption{Probability densities $P(y_{\min})$
of the rescaled minimal spacing  $y_{\min}=s_{\min}/ \langle s_{\min} \rangle$
for tensor products of $\CUE$ random unitary matrices  $\CUE_n \otimes \CUE_n$
for $n=2$ ($\circ$), $n=3$ ($\square$), and $n=8$ ($\triangledown$).
The symbols denote numerical results obtained for $2^{14}$ independent matrices,
solid curve represents the Poissonian distribution,
while dashed line corresponds to eq. (\ref{eq2:Smincuexcue}).
}
\label{fig:xmin1}
\end{figure}

To analyze the distribution of the minimal spacing $P( s_{\rm min})$ for the
tensor products of random unitary matrices it is convenient to introduce an
auxiliary variable $y_{\min}=s_{\rm min}/\langle s_{\rm min} \rangle$.
Probability distribution $P(y_{\rm min})$ is presented in
Fig.~\ref{fig:xmin1} for the $n \times n$ systems with $n=2,3$ and $8$.
Numerical results for $n=2$ agree with an explicit analytical prediction
(\ref{eq2:Smincuexcue}). Due to the tensor product structure of the ensemble
the effect of level repulsion, characteristic of $\CUE$, is washed out.

\begin{figure}[htp]
	\centering
	\includegraphics[width=0.5\textwidth]{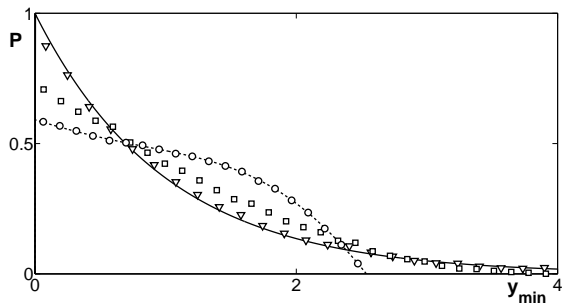}
	\caption{As in Fig.~\ref{fig:xmin1}
    for tensor products of $k$ independent Haar random unitary matrices
   of order two,  ${CUE_2}^{ \otimes k}$
 for $k=2$ ($\circ$), $k=3$ ($\square$), and $k=8$ ($\triangledown$).}
\label{fig:xmin2}
\end{figure}

For larger $n$ the opposite effect of level clustering (large probability at
small values of the minimal spacing) becomes stronger and already for $n=8$
probability distribution can be approximated by the exponential distribution,
$P(y_{\rm min}) =\exp(-y_{\rm min})$, typical of the Poissonian distribution.
A similar transition from distribution (\ref{eq2:Smincuexcue}) to the Poisson
distribution occurs in the case of $k$-qubit systems, as shown in
Fig.~\ref{fig:xmin2}.

\subsection{Maximal spacings for tensor products}

\begin{figure}[htp]
	\centering
	\includegraphics[width=0.5\textwidth]{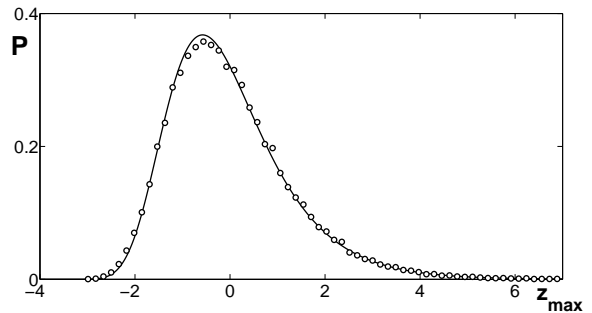}
	\caption{Distribution $P(z_{\max})$ of the deviations  of the rescaled
       maximal spacing from the expected value,
$z_{\max}=\alpha \bigl(s_{\max} - \langle s_{\max} \rangle \bigr)$
with $\alpha = \pi/ \sqrt{ 6  \Var (s_{\max}) }$
for ensemble of  $\CUE_{2^6 \otimes 2^6}$ matrices ($\circ$).
Numerical data obtained out of $2^{16}$ realizations while solid line
denotes the Gumbel distribution (\ref{gumb}).}
\label{fig:xmax1}
\end{figure}

As in section \ref{sec:extremal_statistics:P(S_k)}
we rescale the maximal spacing
$s_{\rm max}$ and analyze the rescaled deviation from the
expectation value

\begin{equation}
\label{zzz}
z_{\max}= \frac{ \pi}{ \sqrt{ 6  \Var (s_{\max}) }}
\bigl(s_{\max} - \langle s_{\max} \rangle\bigr).
\end {equation}
The normalization factor is adjusted to predictions
for the Poissonian process, for which the distribution
of the variable $z$ is  asymptotically described
by the Gumbel distribution,
\begin{equation}
\label{gumb}
P(z) = e^{-(z+\gamma)-e^{-(z+\gamma)}} .
\end{equation}
Recall that $\gamma\approx 0.5772$ denotes Euler's constant,
while the variance of the Gumbel distribution equal to $\pi^2/6$
suggests the convenient prefactor in the definition (\ref{zzz}).
Numerical results on the distributions of the  variable $z_{\max}$
characterizing the distribution of the maximal spacings for the tensor
products corresponding to two qunits and several qubits are presented in
Fig.~\ref{fig:xmin1} and Fig.~\ref{fig:xmax1}, respectively. In the
asymptotic limit of a large matrix size numerical data seem to agree with
predictions (\ref{gumb}) of the Poisson ensemble.

\begin{figure}[htp]
	\centering
	\includegraphics[width=0.5\textwidth]{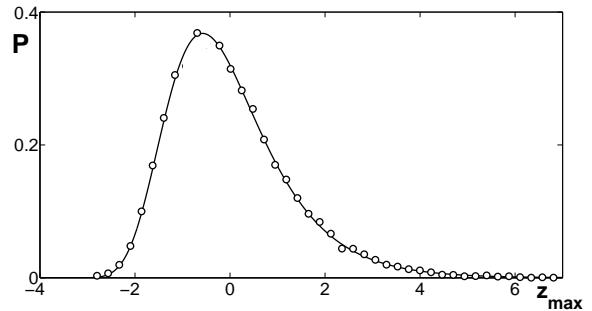}
	\caption{As in Fig.~\ref{fig:xmax1} for
     a sample of $10^5$ realizations of tensor products of $k=22$
     random unitary matrices of order two.}\label{fig:xmax2}
\end{figure}



%

\section{Concluding remarks}\label{sec:remarks}

A significant and spectacular difference between the Poissonian ensemble on
one side and $\COE$ and $\CUE$ on the other, 
concerning the degree of ``repulsion'' between adjacent levels can be
effectively analyzed in terms of distributions of the extremal spacings. We
analyzed the average minimal spacing for several ensembles of random unitary
matrices. Basing on numerical results we propose a general form of the
probability distribution $P(s_{min})$ of the minimal spacing for the standard
ensembles of random unitary matrices. For $\CUE$ this distribution coincides
with the recent result derived by Arous and Bourgade \cite{AB10}, while for
$\COE$ it corresponds to the distributions analyzed for real symmetric
matrices in \cite{CMD07,Fo10}.

The key part of this work concerned tensor products of random unitary
matrices. In the case  of $k$ independent random matrices of order $n$
distributed according to the Haar measure
the tensor product leads asymptotically to a spectrum with Poissonian level
spacing distribution \cite{TSKZZ12,Tk13}. However, we  report here a
different behavior for the statistics of the extreme spacings. Even though
the mean largest spacing $\langle s_{max} \rangle$ can be described by
predictions obtained for the Poisson ensemble of diagonal random unitary
matrices of size $N=n^k$, this is not the case for the mean minimal spacings.

In particular, in the case of $k$ non-interacting qubits,
described by the tensor product $CUE^{\otimes k}$,
the mean minimal spacing $\langle s_{min} \rangle$
displays significant deviations with respect to the
predictions of the Poisson ensemble. In the simplest case
of a two qubit system we have shown that the eigenphases of the tensor product, $CUE_{2\otimes 2}$,
show weaker repulsion than in the case of random CUE matrices of order $N=4$.

Our study leaves several questions open. In particular, numerical results
encourage one to derive an unknown scaling law of the average minimal spacing
 $\langle s_{\min} \rangle$  in the case of $k$-qubit system.
Furthermore, our observations suggesting that
the distributions of the extremal spacing
for ensembles of random matrices corresponding
to two--qunit or $k$--qubit systems are asymptotically governed
by the Poisson and the Gumbel distributions, respectively,
should be confirmed by an analytical proof.

%


\section*{Acknowledgements.}

It is a pleasure to thank L. Erd{\"o}s and O. Zeituni for fruitful remarks
and to P. Forrester for a helpful correspondence.
Financial support by the SFB Transregio-12 project der Deutschen Forschungsgemeinschaft and the grant financed by the Polish National Science Center under the contracts
number DEC-2011/01/M/ST2/00379  (MK,K{\.Z}) as well as Grant number 2011/03/N/ST2/01968 (MS)
 is gratefully acknowledged.

\appendix
\section{Basic properties of the Poisson process}

By a \emph{point process} $\Xi$ on the real half-line $\R_+ = [0,\infty)$
we mean a countable collection of random nonnegative numbers.
For instance, a set $\Xi_U = \{(N/2\pi)\theta_1, \ldots, (N/2\pi)\theta_N\}$ of the rescaled eigenvalues of a random unitary matrix $U$ can be viewed as a point process on $\R_+$.

A key example is a homogeneous \emph{Poisson point process} $\Pi$ on $\R_+$ with a parameter $\lambda > 0$ which is characterized by
\begin{enumerate}[(i)]
	\item for any pairwise disjoint and measurable subsets
   $B_1, \ldots, B_n$ of $\R_+$ the number of points in these subsets
 form independent random variables,

	\item for any measurable subset $B$ of $\R_+$ the number of
points contained inside is described by the Poisson distribution with parameter $\lambda|B|$, where $|B|$ denotes the Lebesgue measure of $B$.
\end{enumerate}
A detailed treatment of this process can be found in a classical monograph \cite{K}.
In this work we set the parameter $\lambda$ to $1$  and call it the standard Poisson point process.

One of the fundamental property of the Poisson process is
that its spacings are independent and are described by exponential distributions.
We read  in \cite{K}
\begin{thm}\label{thm:poissonspacings}
Let $\Pi = \{X_1, X_2, \ldots\}$ be the standard Poisson point process, where the points are labeled so that they do not decrease.
Define its spacings $Y_1, Y_2, \ldots$ by \eqref{eq:defspacingsofPi}. Then the variables $Y_1, Y_2, \ldots$ are independent and identically distributed with density $e^{-y}$, $y>0$.
\end{thm}
Knowing this we are able to examine the asymptotics of the extreme gaps $Y_{\min}=\min_{j \leq N} Y_j$ and $Y_{\max}=\max_{j \leq N} Y_j$.
\begin{thm}\label{thm:asymptofextremepoisson}
Let $Y_1, Y_2, \ldots$ be a sequence of random variables which are independent identically distributed with density $P(y)=e^{-y}$ for $y > 0$. Then,
\begin{equation}\label{eq:meanY}
\begin{split}
	\langle Y_{\min} \rangle &= \langle \min_{j \leq N} Y_j \rangle= 1/N, \\
	\langle Y_{\max} \rangle &= \langle \max_{j \leq N} Y_j \rangle = \sum_{k=1}^N 1/k \sim \ln N.
\end{split}
\end{equation}

If we rescale the variables to set the mean to unity, $y=NY$, asymptotically they behave exponentially and concentrate respectively,
\begin{equation}\label{eq:Yconvergence}
	N Y_{\min} \overset{d}{\longrightarrow}  e^{-y}\1_{\{y > 0\}},
\end{equation}
\begin{equation}\label{eq:Yconvergence1}
	Y/\langle Y_{\min} \rangle \overset{d}{\longrightarrow} 1,
\end{equation}
where $\overset{d}{\longrightarrow}$ denotes the convergence in distribution.

Furthermore, the fluctuations of $Y/\langle Y_{\min} \rangle$ around $1$ are governed at the scale $ \langle \max_{j \leq N} Y_j \rangle \sim \ln N$ by the Gumbel distribution,
\begin{equation}\label{eq:Yfluctuations}
	Y - \langle Y_{\min} \rangle \overset{d}{\longrightarrow} P(z)= e^{-(z+\gamma)-e^{-(z+\gamma)}},
\end{equation}
where $\gamma := \lim_{n \to \infty} \left( \sum_{k=1}^n 1/k - \ln n \right) \approx 0.5772$ is Euler's constant.
\end{thm}

Given the fact that the distribution functions are easily calculable,
\begin{align*}
	\p{\min_{j \leq N} Y_j > t} &= e^{-Nt}, \quad t > 0, \\
	\p{\max_{j \leq N} Y_j \leq t} &= (1 - e^{-t})^N, \quad t > 0,
\end{align*}
theorem 2 can be proved by a direct computation.


\end{document}